\documentclass[twocolumn,pre,aps,10pt]{revtex4-1}

\usepackage{amsfonts,amssymb}
\usepackage[sumlimits,intlimits]{amsmath}
\usepackage{graphics}
\usepackage{graphicx}
\usepackage{epsfig}
\usepackage{mathrsfs}
\usepackage{verbatim}
\usepackage{hyperref}    
\usepackage{bm}          

\newcommand{\be}{\begin{equation}}
\newcommand{\ee}{\end{equation}}

\begin{document}

\title{Coulomb gap in the one-particle density of states in three-dimensional systems with localized electrons}

\date{\today}

\author{A. L. Efros}
\affiliation{Department of Physics and Astronomy, University of Utah, Salt Lake City, Utah 84112}
\author{Brian Skinner}
\author{B. I. Shklovskii}
\affiliation{Fine Theoretical Physics Institute, University of Minnesota, Minneapolis, Minnesota 55455}

\begin{abstract}

The one-particle density of states (1P-DOS) in a system with
localized electron states vanishes at the Fermi level due to the
Coulomb interaction between electrons. Derivation of the Coulomb
gap uses stability criteria of the ground state. The simplest
criterion is based on the excitonic interaction of an electron and
a hole and leads to a quadratic 1P-DOS in the
three-dimensional (3D) case. In 3D, higher stability criteria, including two or more electrons, were predicted to exponentially deplete the 1P-DOS
at energies close enough to the Fermi level. In this paper we show
that there is a range of intermediate energies where this
depletion is strongly compensated by the excitonic interaction
between single-particle excitations, so that the crossover
from quadratic to exponential behavior of the 1P-DOS is retarded.
This is one of the reasons why such exponential depletion was never seen in computer simulations.

\end{abstract}
\maketitle

\section{Introduction} \label{sec:intro}

The study of localized, interacting electrons in a disordered
system has been a source of interesting physics for nearly half a
century.  The canonical example of such a system is a lightly
doped, compensated, $n$-type semiconductor, where electrons become
localized on donor sites~\cite{SE1984}. The study of
electron-electron interactions in the localized regime was
initiated by Pollak~\cite{Pollak} and
Srinivasan~\cite{Srinivasan}. Efros and
Shklovskii~\cite{ES1975,Efros} later argued that the single
particle density of states (1P-DOS) tends to zero at the Fermi
level as a result of the long-range part of the Coulomb
interaction between electrons. They proposed the following
universal soft gap in the 1P-DOS near the Fermi level at
temperature $T = 0$, which is called the Coulomb gap and depends
only on the dimensionality $D$ of the system: \be g(\epsilon)=
\frac{2}{\pi e^4}|\epsilon|  \textrm{ for $D=2$} \label{eq:g2D}
\ee and \be g(\epsilon)= \frac{3}{\pi e^6}\epsilon^2  \textrm{ for
$D=3$.} \label{eq:g3D} \ee Here the reference point for the energy
$\epsilon$ is the Fermi level and $e^2$ denotes the square of the
electron charge divided by the dielectric constant. Eqs.\
(\ref{eq:g2D}) and (\ref{eq:g3D}) were derived for the case when
the bare DOS, which is the DOS without Coulomb interactions, has a
non-zero value at $\epsilon=0$.

The primary method for quantitative, theoretical study of the
Coulomb gap is through computer simulations
\cite{Baran1979,Levin,Lee,Mobius,Li,Zimanyi,Mobius3,Palassini,Mobius2},
which mostly confirm the above results for the 1P-DOS. The most
important experimental manifestation of the Coulomb gap is its
effect on the variable range hopping conductivity $\sigma$, which
as a consequence of the Coulomb gap obeys the law~\cite{ES1975},
\be \sigma = \sigma_0 \exp[-(T_0/T)^{1/2}], \label{eq:ESlaw} \ee
observed in dozens of experimental works. Here $\sigma_0$ is a
constant and $T_0 = \beta_D e^2/\xi$, where $\xi$ is the
localization length and $\beta_D$ is a numerical constant that
depends on the dimensionality: $\beta_3 \approx 2.8$ \cite{SE1984}
while $\beta_2 \approx 6$ \cite{Nguyen,tsig2002}.

Eqs.\  (\ref{eq:g2D}) and (\ref{eq:g3D}) are derived from the
requirement that the ground state be stable with respect to the
transfer of a single electron from an occupied to an empty state.
In this introduction we will just mention that this depletion of
the bare 1P-DOS is a result of the excitonic attraction of
electrons and holes near the Fermi level, which eliminates
electrons and holes with small $|\epsilon|$ that are close in
space, pushing them to higher energies. In other words, one can
say that electrons and holes that remain in the ground state and
contribute to the 1P-DOS ``intimidate" each other.

While Eqs.\ (\ref{eq:g2D}) and (\ref{eq:g3D}) reflect the effect
of this first-order stability criterion, a more challenging
question is how the 1P-DOS is affected by higher-order stability
criteria, in which two or more electrons are transferred
simultaneously.  It has been shown~\cite{Efros} that such criteria
do not significantly modify the linear gap in the 2D case [Eq.\
(\ref{eq:g2D})], but they should modify the quadratic gap in the
3D case [Eq.\ (\ref{eq:g3D})]. Previous
authors~\cite{Efros,Baran1980} have suggested that compact dipole
excitations intimidate single particle excitations, thereby making
the 1P-DOS exponentially small in the limit of small enough
energies. It was also understood that this effect does not modify
the law of variable-range hopping conductivity [Eq.\
(\ref{eq:ESlaw})] because when dipole excitations play an
important role hopping conductivity occurs by multi-electron
excitations called electronic polarons, for which the DOS always
obeys Eq.\ (\ref{eq:g3D}) \cite{Efros,SE1984,ES1985}.  The
depleted 1P-DOS could in principle be seen in tunneling
experiments~\cite{MarkLee}.

However, such strong, exponential depletion of the 1P-DOS was never seen in computer experiments~\cite{Baran1979,Levin,Lee,Mobius,Li,Zimanyi,Mobius3,Palassini,
Mobius2}.  For large simulated samples this is the case because
computer simulations are not able to reach the ground state of the system
or cannot enforce a sufficiently large number of higher-order stability criteria.
In this paper we show
that even for samples where the ground state can be reached the
crossover from a quadratic to an exponentially small 1P-DOS with
decreasing energy $|\epsilon|$ is strongly retarded. The reason is
that the survival probability
$P(\epsilon)$ of single particle excitations in the presence of
intimidating dipoles remains a weaker function of $\epsilon$
than Eq.\ (\ref{eq:g3D}) over a wide range of energies. The crucial idea of this paper is
that the bare DOS multiplied by $P(\epsilon)$ can be considered as
an effective new bare DOS. We show that because of the remarkable
stability of $g(\epsilon)$ with respect to changes in the bare DOS,
such modification produces only a small effect on the 1P-DOS over
a substantial range of intermediate energies.

Finally, at even smaller energies, where the exponential effect of
dipoles reduces the modified bare DOS so strongly that the mutual
electron-hole intimidation plays a weak role, the 1P-DOS
decreases exponentially and is proportional to $P(\epsilon)$.
At present, computer simulations are not able to find the
ground state in systems that are large enough to explore such
small energies.

The remainder of this paper is organized as follows. In Sec.\
\ref{sec:Exciton} we present the simple model Hamiltonian that is
used in most numerical studies of the Coulomb gap and we discuss
the role of mutual intimidation of single particle excitations,
including its description by the self-consistent equation
(SCE)~\cite{Efros,ES1985}. In Sec.\ \ref{sec:dipoles} we
recapitulate old theoretical results related to dipole excitations
and their mutual intimidation~\cite{Baran1980}. We then discuss
the depletion of the bare DOS by surviving dipole
excitations~\cite{Efros,Baran1980} and we use this depleted DOS as a
bare DOS to solve the SCE for $g(\epsilon)$. In Sec.\
\ref{sec:Powers} we study analytically the stability of the
Coulomb gap for hypothetical cases where the bare DOS follows a
power law, and we confirm all qualitative conclusions obtained
numerically in Sec.\ \ref{sec:dipoles}.

\section{Excitonic effect on single particle density of states} \label{sec:Exciton}

In order to remind the reader of the derivation of Eqs.\
(\ref{eq:g2D}) and (\ref{eq:g3D}), we start from the model
Hamiltonian~\cite{Efros}: \be H = \sum_i \phi_i n_i + \frac{1}{2}
\sum_{i,j} \frac{e^2}{r_{ij}} (n_i-1/2)(n_j-1/2).
\label{eq:Ehamil} \ee The electrons described by this Hamiltonian
are assumed to occupy sites on a cubic lattice with lattice
constant $l$.  $n_i = 0,1$ is the occupation number of site $i$
and $r_{ij}$ is the distance between sites $i$ and $j$. The
quenched random site energies $\phi_i$ are distributed uniformly
within the interval $[-Ae^2/l, Ae^2/l]$, where $A \gg 1 $ is some
positive number, so that the bare DOS created by the random site
energies is the ``box function" $g_0(\epsilon) = g_{00} \Theta[A -
|\epsilon|/(e^2/l)]$. Here, $g_{00} = (2Ae^2 l^2)^{-1}$ and
$\Theta[x]$ denotes the Heaviside step function. In order to make
the system neutral each site is given a positive background charge
$e/2$. Due to electron-hole symmetry, the chemical potential $\mu
= 0$. The single particle energy at site $i$ is given by \be
\epsilon_i = \phi_i + \sum_{j} \frac{e^2}{r_{ij}} (n_i-1/2).
\label{eq:Energy} \ee

At zero temperature (in the ground state of the system) all states
with $\epsilon_i < 0$ are occupied and all states with $\epsilon_i
> 0$ are empty, since the ground state must be stable with respect
to transfer of an electron from an occupied site to infinity or
from infinity to an empty site.  This condition is only the first
stability criterion of the ground state. The second stability
criterion considers the transfer of one electron from some site
$i$ that is occupied in the ground state to a site $j$ that is
vacant in the ground state. The change in $H$ resulting from such
a transfer, 
\be 
\Omega_{ij} = \epsilon_j - \epsilon_i -
\frac{e^2}{r_{ij}}, \label{eq:Exciton} 
\ee
must be positive for the stability of the ground state.  The last
term in Eq.\ (\ref{eq:Exciton}) reflects the simple fact that the
ground state energy $\epsilon_j$ includes the potential of the
transferred electron, which initially was at site $i$.

One can see the origin of the Coulomb gap from Eq.\
(\ref{eq:Exciton}). Since $\epsilon_j
> 0$ and $\epsilon_i < 0$, the first two terms give a positive
contribution to $\Omega_{ij}$, while the third term is negative.
Thus, any two sites $i,j$ with energy close to the Fermi level
should be separated in space to keep $\Omega_{ij} > 0$.  To see
how this produces the Coulomb gap, consider sites whose energies
fall in a narrow band  $(-\epsilon/2, \epsilon/2)$ around the
Fermi level.  According to Eq.\ (\ref{eq:Exciton}), any two sites
in this band with energies on opposite sides of the Fermi level
must be separated by a distance $r_{ij}$ not less than
$e^2/\epsilon$. Therefore the concentration of such sites
$n(\epsilon)$ cannot exceed $(\epsilon/e^2)^D$. Thus, the 1P-DOS $g(\epsilon) =
dn(\epsilon)/d\epsilon$ must vanish when $\epsilon$ tends to zero
at least as fast as $\epsilon^{D-1}$.

In this way we arrive at a simple upper bound for the 1P-DOS.
Applying the principle of the  micro-canonical distribution to the
disordered system we find that all states except the forbidden ones
are homogeneously distributed in phase space. Thus, the bound $g(\epsilon) \propto \epsilon^{D-1}$ is
an exact one if we enforce only the criterion $\Omega_{ij} > 0$.
Applying all other stability criteria can only reduce the 1P-DOS.

The numerical coefficients in Eqs.\ (\ref{eq:g2D}) and
(\ref{eq:g3D}) were later found by solving the SCE for
$g(\epsilon)$, which is designed to describe mutual intimidation
of single particle excitations.  In 3D this equation has the
form~\cite{Efros,ES1985} \be g(\epsilon) = g_0(\epsilon) \exp
\left[ -\frac{2 \pi}{3} e^6 \int_0^\infty \frac{g(\epsilon')
d\epsilon'}{(|\epsilon| + \epsilon ' )^3 } \right],
\label{eq:SCE3D} \ee where $g_0(\epsilon)$ is the bare DOS.

The asymptotic solution of Eq.\ (\ref{eq:SCE3D}) at small energies
$\epsilon \ll \Delta$ leads to Eq.\ (\ref{eq:g3D})~\cite{Efros}.
Here $\Delta$ is the width of the Coulomb gap, defined as the
energy $\epsilon$ at which Eq.\ (\ref{eq:g3D}) matches $g_{00}$,
so that $\Delta = (e^2/l)(\pi/6A)^{1/2} \ll A(e^2/l) $. Solution
of the 2D version of the SCE at small energies $\epsilon \ll
\Delta$ leads to  Eq.\ (\ref{eq:g2D}). Remarkably, these
asymptotic solutions do not depend on $A$ or $l$ and in this sense
are universal. It should be noted that Eq.\ (\ref{eq:SCE3D}) does
not describe the corresponding increase in high energy states that
accompanies the depletion of low energy states. In this sense its
solution $g(\epsilon)$ is not exactly normalized to contain the
same total number of states as $g_0(\epsilon)$, but this should
not be very important for $A \gg 1$ .

At only moderately small energies, where $\epsilon/\Delta$ is not
asymptotically small, the integral equation of Eq.\
(\ref{eq:SCE3D}) has no known analytical solution.  Nonetheless,
one can find numerical solutions to the SCE by using a simple
iteration procedure, wherein an initial guess is made for the
function $g(\epsilon)$ [say, that of Eq.\ (\ref{eq:g3D})] and then
inserted into the right-hand side of Eq.\ (\ref{eq:SCE3D}) to
produce a new estimate for $g(\epsilon)$.  This iteration can be
repeated an arbitrary number of times until a convergent solution
is found for the function $g(\epsilon)$ over the desired range of
energies~\cite{convergencefootnote}.

The result of this procedure is shown for $A=2$ in Fig.\
\ref{fig:solution} by the thin red line. At small energies this
solution approaches its asymptotic form [Eq.\ (\ref{eq:g3D}),
shown by the dotted black line]. In the dimensionless
units of this plot the solution for $A=4$ (not shown)
coincides almost exactly with the solution for $A=2$.

\section{Dipole excitations and their interaction with single particles excitations} \label{sec:dipoles}

Until now we have taken into account only the conditions that $H$
be minimized with respect to the change of one or two occupation
numbers. An analysis of the role of stability criteria with
respect to the change of three or more occupation numbers shows
that in the 2D case the single-electron approach, used above, is
good, while in the 3D case the physics is more
complicated~\cite{Efros}.

It is convenient to talk about this problem by introducing a
dipole branch of excitations. For a given electron transition from
a site that is occupied in the ground state to an empty site, the
energy is given by  Eq.\ (\ref{eq:Exciton}).  The result of such a
transition is the formation of an ``electron-hole pair".  If the
energy $\Omega_{ij}$ of this transition satisfies $\Omega_{ij} \gg
e^2/r_{ij}$, then the pair can be thought of as two independent
single-particle excitations: an extra electron on site $j$ and an
extra hole on site $i$. If $\Omega_{ij}\ll e^2/r_{ij}$, on the
other hand, the pair constitutes a strongly bound small
dipole~\cite{ES1975,Efros,Baran1980}. The typical arm of such
dipoles is $r_0 = e^2/\Delta$.  At small $\Omega$ these dipoles
are spatially separated from each other and do not participate in
dc conductivity.  They do contribute, however, to the ac
conductivity and the low temperature thermodynamics~\cite{ES1975,Efros,Baran1980}.

In the first approximation the interaction between these dipole
excitations was assumed to be weak~\cite{ES1975,Efros} and their
density of states (2P-DOS) $\Phi(\Omega)$ was predicted to be
constant at small energies $\Omega$ and given by $\Phi(\Omega)=
g_{00} $.  A more careful consideration \cite{Baran1980,ES1985}
shows that dipole-dipole interaction leads to some mutual
intimidation, resulting in the logarithmic depletion of
$\Phi(\Omega)$.  Specifically, $\Phi(\Omega) =
g_{00}/[\ln(\Delta/\Omega)]^{1/2}$. At the same time, the
remaining soft dipoles have dipole arms that are shorter than $r_0$,
with an average arm $r_0/[\ln(\Delta/\Omega)]^{1/4}$. This happens
because shorter dipoles interact more weakly and therefore survive
with larger probability.

\begin{figure}[htb]
\centering
\includegraphics[width=0.45 \textwidth]{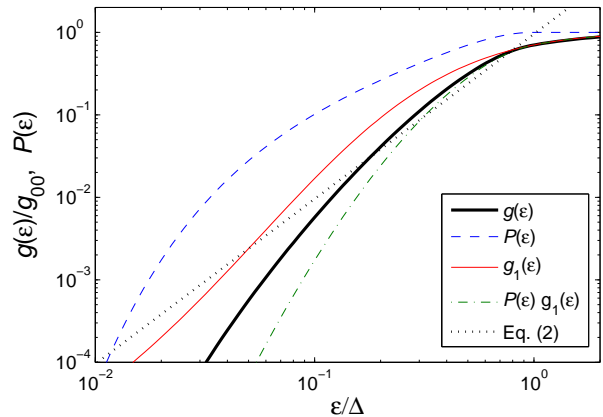}
\caption{(Color online) The 1P-DOS $g(\epsilon)$ (thick
black curve) and the probability $P(\epsilon)$ for a single particle
excitation to survive interaction with dipole excitations (blue,
dashed curve). The thin red curve, $g_1(\epsilon)$, is calculated
from  Eq.\ (\ref{eq:SCE3D}), the green dashed-dotted curve is the
product of this solution with $P(\epsilon)$, and the black dotted
line is the asymptotic result of Eq.\ (\ref{eq:g3D}).  Both
$g(\epsilon)$ and $g_1(\epsilon)$ are calculated using $A = 2$.}
\label{fig:solution}
\end{figure}

Let us now describe the interaction between dipole excitations and
single particle excitations.  We first note that dipole
excitations are more abundant than single particle excitations due
to the strong depletion of the 1P-DOS at the Fermi level. Thus, it
is natural to think that after finding the 2P-DOS one can use it
to study the role of dipoles in intimidating single-particle
excitations. Following this logic Efros~\cite{Efros} noticed that
in 3D a dipole with the proper orientation and proximity can
intimidate a single particle excitation. Using the constant 2P-DOS
$\Phi(\Omega) = g_{00}$, Efros showed that the probability that a
single particle will survive this interaction is  given by
$P(\epsilon) = \exp[-(\Delta/\epsilon)^{1/2}]$. An improved
understanding of the effects of dipole-dipole
interactions~\cite{Baran1980} leads to a logarithmic increase of
this probability. A calculation along the lines of
Ref.~\onlinecite{Baran1980} shows that a more accurate expression
for the survival probability is \be P_1(\epsilon) = \exp \left[ -
\gamma^{1/2}\left( \frac{\Delta}{\epsilon} \right)^{1/2}
\ln^{-7/8} \left( \frac{\Delta}{\epsilon} \right) \right],
\label{eq:OnedipoledepletionP} \ee where $\gamma = 1.5$ is a
numerical constant~\cite{Baran1980}. Both the logarithmic
reduction of $\Phi(\Omega)$ and the logarithmic reduction of the
size of dipoles discussed above contribute to the logarithmic
factor in Eq.\ (\ref{eq:OnedipoledepletionP}) that increases the
probability $P_{1}(\epsilon)$. The subscript $1$ in
$P_1(\epsilon)$ emphasizes that it describes the effect of intimidation by a  single dipole.

Later in Ref.~\onlinecite{Baran1980}, the authors realized that
collective intimidation by many dipoles surrounding a one particle
excitation is stronger than intimidation by a single dipole. This
can be seen by considering that around an empty site (a positive
single particle excitation) there is a certain probability to find
a cloud of many dipole moments oriented away from the hole. Such
dipoles can re-polarize (move their electron away from hole) when
an electron is brought to the empty site from a low-energy
occupied site.  This process lowers the system's total energy.  To
guarantee that the original state is not vulnerable to such an
event, and, therefore, is the true ground state, one has to
exclude such a polarization atmosphere.

The probability of a hole or electron surviving in the ground
state was calculated in Ref.~\onlinecite{Baran1980}: \be
P(\epsilon) = \exp \left[-\gamma \left(
\frac{\Delta}{\epsilon}\right)\ln^{-7/4}
\left(\frac{\Delta}{\epsilon}\right)\right].
\label{eq:ManydipoledepletionP} \ee

Remarkably, the argument of the exponential in $P(\epsilon)$ is
exactly the square of the argument of the exponential of
$P_{1}(\epsilon)$. Thus, in the interesting case where the latter
argument is larger than unity, the intimidation of single particle
excitations is determined by the collective effect result of Eq.\
(\ref{eq:ManydipoledepletionP}).

Eq.\ (\ref{eq:ManydipoledepletionP}) is of course derived in the
limit that $\ln (\Delta/\epsilon) \gg 1$.  Attempting to use it at
$\Delta/\epsilon \sim 1$ leads to nonphysical divergence, since
$\ln (\Delta/\epsilon)$ tends to zero.  Instead, the correct
behavior for $P(\epsilon)$ is that it should approach unity at
$\epsilon/\Delta \geq 1$.
One can model such behavior while still maintaining the correct
small energy asymptotic form in Eq.\
(\ref{eq:ManydipoledepletionP}) by introducing a multiplicative
``crossover function" $f(\epsilon/\Delta)$ into the exponent of
Eq.\ (\ref{eq:ManydipoledepletionP}): 
\be 
P(\epsilon) = \exp
\left[-\gamma \left( \frac{\Delta}{\epsilon}\right)\ln^{-7/4}
\left(\frac{\Delta}{\epsilon}\right)
f\left(\frac{\epsilon}{\Delta} \right)\right].
\label{eq:ManydipoledepletionPA} 
\ee 
The function $f(x)$ should
have the properties $f(x \rightarrow 0) = 1$ and $f(x > 1) = 0$.
One should also ensure that $f(x)$ approaches zero at $x = 1$
faster than $(x-1)^{7/4}$, so that the divergent behavior of Eq.\
(\ref{eq:ManydipoledepletionP}) at $\epsilon/\Delta = 1$ is
eliminated.  One obvious choice for $f(x)$ is \be f(x) = (1 -
x)^\eta \Theta(1 - x) \ee with $\eta > 7/4$.  In the numeric
results presented in Fig.\ \ref{fig:solution} we use $\eta = 4$,
which produces a smooth crossover from the asymptotic behavior of
Eq.\ (\ref{eq:ManydipoledepletionP}) to $P(\epsilon) = 1$ at
$\epsilon > \Delta$.

We can now consider how the 1P-DOS is affected by the elimination
of single particle states, as given by the small
survival probability  in Eq.\ (\ref{eq:ManydipoledepletionPA}).
Asymptotically at very small $\epsilon$, $P(\epsilon)$ goes to
zero faster than $\epsilon^2$ and should play a dominating role in
the 1P-DOS.  This is why the fact that numerical data for
$g(\epsilon)$ in 3D are close to Eq.\ (\ref{eq:g3D}) looked
puzzling~\cite{SE1984,ES1985}.

We would like to argue here that the dominance of  Eq.\
(\ref{eq:ManydipoledepletionPA}) requires very small $\epsilon$.
One can notice that around $\epsilon/\Delta = 0.3$ the probability
$P(\epsilon)$ still has the strength of the first power of
$\epsilon/\Delta$. It reaches the strength of
$(\epsilon/\Delta)^2$ only at $\epsilon/\Delta \approx 0.1$. This
is seen in Fig.\ \ref{fig:solution}, where the probability
$P(\epsilon)$ is plotted by the dashed blue line.

Still, one may think that the effect of mutual intimidation of
single particle excitations and the effect of intimidation of
single particles by dipole excitations are multiplicative, so that
to obtain $g(\epsilon)$ one should multiply the solution of Eq.\
(\ref{eq:SCE3D}) for the bare ``box" DOS, which we dub
$g_1(\epsilon)$ (thin red line), by $P(\epsilon)$ from Eq.\
(\ref{eq:ManydipoledepletionPA}). This product is shown by the
dashed-dotted green curve in Fig.\ \ref{fig:solution}.

We argue that instead of directly multiplying the final result of
Eq.\ (\ref{eq:SCE3D}) by $P(\epsilon)$, one should think that the
role of $P(\epsilon)$ is to modify the bare DOS $g_0(\epsilon)$.
Thus, to find $g(\epsilon)$ one should replace $g_0(\epsilon)$ in
Eq.\ (\ref{eq:SCE3D}) by $g_0(\epsilon)P(\epsilon)$ and solve this
new self-consistent equation for $g(\epsilon)$. The result of this
calculation is shown by the thick black line in Fig.\
\ref{fig:solution}, as determined by the numerical iteration
procedure described in Sec.\ \ref{sec:Exciton}. 

One can see that at $\epsilon/\Delta > 0.1$ the 1P-DOS stays much closer to Eq.\
(\ref{eq:SCE3D}) than the product $P(\epsilon) g_1(\epsilon)$.
Thus, in the range of energies  $\epsilon
> 0.1 \Delta$, the solution of Eq.\ (\ref{eq:SCE3D}) is very
stable with respect to modification of the bare density of states.
The mechanism of this stability is as follows.  When
$g_0(\epsilon) P(\epsilon)$ becomes moderately small, the
exponential factor on the right-hand side of Eq.\ (\ref{eq:SCE3D})
grows sharply, supporting an almost unchanged value for
$g(\epsilon)$. In other words, any attempt at moderate reduction
of $g(\epsilon)$ by dipole intimidation is met by resistance from
the compensating effect of weakening of the intimidation by
single-particle excitations. This is an extension of the
universality of the small energy solution of Eq.\ (\ref{eq:SCE3D})
with respect to varying the bare ``box" density of states
$g_{00}$~\cite{SE1984,ES1985}.

Returning to Fig.\ \ref{fig:solution}, we note that at $\epsilon \ll
0.1$, where the probability $P(\epsilon)$ changes substantially
faster than $(\epsilon/\Delta)^2$, our result for $g(\epsilon)$
follows the exponentially small $P(\epsilon)$. At present, however,
this is a purely theoretical range, since computer simulations can
find the ground state at $A\sim 3$ only for lattices of size $10
\times 10 \times 10$ or smaller~\cite{PalassiniFuture}. For such
small samples, finite size effects~\cite{Baran1979} limit the range
over which one can determine the 1P-DOS of infinite systems to
$\epsilon/\Delta \gtrsim 0.3$~\cite{PalassiniFuture}. At these
energies the value of $g(\epsilon)$ in the ground state follows Eq.\
(\ref{eq:g3D})~\cite{PalassiniFuture}, in agreement with our
prediction in Fig.\ \ref{fig:solution} (the thick black curve).

\section{Stability of the Coulomb gap: Analytical solutions} \label{sec:Powers}

In the previous section we dealt with the stability of the 1P-DOS Coulomb gap with respect to moderate depletion of the bare DOS
$g_0(\epsilon)$ by solving the SCE numerically. In
this section we would like to illustrate this stability for an
analytically solvable model. We consider this model only in its 3D version, but the two dimensional version is similar.

Assume that the bare ``box" DOS $g_0(\epsilon) = g_{00}
\Theta[A - |\epsilon|/(e^2/l)]$ used above is modified at small
energies $\epsilon < \Delta$ by the power law factor
\be
g_0=g_{00}(\epsilon/\Delta)^\alpha,
\label{go}
\ee
where $\alpha \geq 0$. In order to find out how this changes the low energy solution of Eq.\ (\ref{eq:SCE3D}), we write it in the differential form
\begin{equation}\label{dif}
\frac{d\ln g}{d\epsilon}=\frac{d\ln g_0}{d\epsilon}+ 2\pi
e^6\int_0^\infty\frac{g(\epsilon')d\epsilon'}{(\epsilon +\epsilon')^{4}}.
\end{equation}

At small values of $\epsilon$ and $\alpha < 2$ the solution of this
equation is
\be
g(\epsilon)=\beta\epsilon^2/e^6,
\label{g}
\ee
where $\beta$ is a numerical factor. Substituting Eq.\ (\ref{g})
into Eq.\ (\ref{dif}) one gets
\begin{equation}\label{be}
\beta= \frac{3(2-\alpha)}{2\pi}.
\end{equation}

At $\alpha=0$ we are back to the bare ``box" $g_0(\epsilon)$ and Eq.
(\ref{g}) coincides with Eq.\ (\ref{eq:g3D}). One can see that even
at $0 < \alpha < 2$ the density of states is still quadratic and
the only effect of the depletion of the bare DOS in Eq.\ (\ref{go}) is to reduce the numerical coefficient $\beta$. In this way at small energies the solution $g(\epsilon)$ is remarkably stable to the strong change in $g_0$.
The reason for this stability is that at $0 < \alpha < 2$ the number of available sites in a small energy band near $\epsilon = 0$ exceeds the number of electrons that is permitted by their interaction.  Of course, the electrons cannot arrange themselves as comfortably as in the case of constant $g_0$, and this fact is reflected in the reduction of the coefficient $\beta$.

On the other hand, when $\alpha > 2$ the number of places for
electrons in a small energy band near $\epsilon=0$ is less than
the number of electrons permitted by their interaction. Therefore,
$g(\epsilon)$ has the same energy dependence as $g_0(\epsilon)$.
This can be seen formally from Eq. (\ref{eq:SCE3D}).
Substitution of  $g_0(\epsilon)$ into the integral term  with $\alpha > 2$ allows one to set $\epsilon = 0$ inside the integral at $\epsilon/\Delta \ll 1$. Then at small
$\epsilon$ one gets $g(\epsilon) = B g_0(\epsilon)$, where $B$ is a constant given by
\begin{equation}\label{B}
B=\exp \left[ -\frac{2 \pi}{3} e^6 \int_0^\infty
\frac{g_0(\epsilon') d\epsilon'}{ (\epsilon ' )^3 } \right].
\end{equation}
Here we have used the fact that $g_0(\epsilon)$ is zero at
$\epsilon > A (e^2/l)$, so that the integral in Eq.\ (\ref{B}) converges.

In the critical case, $\alpha = 2$, one can show that the asymptotic solution at
$\epsilon \rightarrow 0$ is
\begin{equation}\label{alpha2}
g(\epsilon) = \frac{3}{2\pi e^{6}}\frac{\epsilon^{2}}{\ln(\Delta/\epsilon) + h(\Delta/\epsilon)},
\end{equation}
where $h(\Delta/\epsilon)$  is a small correction to $ \ln(\Delta/\epsilon)$.  We were not able to find this correction analytically, but numerical solutions to the SCE suggest that
$h(\Delta/\epsilon) \simeq 7.3$ at $\epsilon/\Delta \ll 10^{-3}$.

The analytical solutions of this section confirm our
numerical experience from the previous section.  Namely, that when the bare density of states $g_0(\epsilon)$ at small energies is larger than the one given by Eq.\ (\ref{eq:g3D}), the solution of Eq.\ (\ref{eq:SCE3D}) is very close to Eq.\ (\ref{eq:g3D}). On the other hand, when $g_0(\epsilon)$ is
smaller than Eq.\ (\ref{eq:g3D}) at small energies the density of states $g(\epsilon)$ is close to $g_0(\epsilon)$.

We are grateful to M. Goethe and M. Palassini for making available
to us their yet unpublished data on the 1P-DOS $g(\epsilon)$ and
2P-DOS $\Phi(\Omega)$; their data stimulated this work. We also
thank them and A.\ M\"{o}bius for their comments on the draft of this work.  We acknowledge the hospitality of the KITP, where this work was started. B.\ S.\ acknowledges financial support from the NSF.


\begin{thebibliography}{99}
\bibitem{convergencefootnote}
In fact, the iterative numerical solution to Eq.\ (\ref{eq:SCE3D})
converges only very slowly at $\epsilon/\Delta < 10^{-2}$, with
the solution oscillating as a function of iteration number. The
convergence of the solution can be greatly sped up if one introduces a relaxation procedure, such that each new solution for $g(\epsilon)$ is
averaged with the previous one using a weighted arithmetic or
geometric mean.  All numeric solutions presented in this paper
use this relaxation procedure, and they were found to be completely independent of the nature of the relaxation process and of the choice of initial guess for
$g(\epsilon)$.
\end{thebibliography}
\end{document}